\documentclass[preprint]{aastex}
\shorttitle{Cosmic-ray P \& He spectra measured with BESS}
\shortauthors{Sanuki et al.}

\begin{document}

\title{Precise Measurement of Cosmic-Ray Proton and Helium Spectra \\
  with the BESS Spectrometer}

\author{
T.~Sanuki\altaffilmark{1},
M.~Motoki\altaffilmark{2},
H.~Matsumoto\altaffilmark{1},
E.~S.~Seo\altaffilmark{5},
J.~Z.~Wang\altaffilmark{5},
K.~Abe\altaffilmark{1},
K.~Anraku\altaffilmark{1},
Y.~Asaoka\altaffilmark{1},
M.~Fujikawa\altaffilmark{1},
M.~Imori\altaffilmark{1},
T.~Maeno\altaffilmark{1},
Y.~Makida\altaffilmark{2},
N.~Matsui\altaffilmark{1},
H.~Matsunaga\altaffilmark{1},
J.~Mitchell\altaffilmark{4},
T.~Mitsui\altaffilmark{3},
A.~Moiseev\altaffilmark{4},
J.~Nishimura\altaffilmark{1},
M.~Nozaki\altaffilmark{3},
S.~Orito\altaffilmark{1},
J.~Ormes\altaffilmark{4},
T.~Saeki\altaffilmark{1},
M.~Sasaki\altaffilmark{3},
Y.~Shikaze\altaffilmark{1},
T.~Sonoda\altaffilmark{1},
R.~Streitmatter\altaffilmark{4},
J.~Suzuki\altaffilmark{2},
K.~Tanaka\altaffilmark{2},
I.~Ueda\altaffilmark{1},
N.~Yajima\altaffilmark{6},
T.~Yamagami\altaffilmark{6},
A.~Yamamoto\altaffilmark{2},
T.~Yoshida\altaffilmark{2},
and
K.~Yoshimura\altaffilmark{1}
}

\altaffiltext{1}{Department of Physics, Faculty of Science,
 University of Tokyo,
 Bunkyo, Tokyo 113-0033,
 Japan;
 sanuki@icepp.s.u-tokyo.ac.jp}
\altaffiltext{2}{High Energy Accelerator Research Organization (KEK),
 Tsukuba, Ibaraki 305-0801,
 Japan}
\altaffiltext{3}{Kobe University,
 Kobe, Hyogo 657-8501,
 Japan}
\altaffiltext{4}{National Aeronautics and Space Administration /
 Goddard Space Flight Center,
 Greenbelt, MD 20771,
 USA}
\altaffiltext{5}{University of Maryland,
 College Park, MD 20742,
 USA}
\altaffiltext{6}{The Institute of Space and Astronautical Science,
 Sagamihara, Kanagawa 229-8510,
 Japan}

\begin{abstract}
We report cosmic-ray proton and helium spectra
 in energy ranges of 1 to 120~GeV and 1 to 54~GeV/nucleon, respectively,
 measured by a balloon flight of the BESS spectrometer in 1998. 
The magnetic-rigidity of the cosmic-rays was reliably determined
 by highly precise measurement of the circular track
 in a uniform solenoidal magnetic field of 1~Tesla. 
Those spectra were determined
 within overall uncertainties of $\pm 5$~\% for protons
 and $\pm 10$~\% for helium nuclei
 including statistical and systematic errors.
\end{abstract}


\section{INTRODUCTION}

The absolute fluxes and spectra
 of primary cosmic-ray protons and helium nuclei
 are fundamental information as references in cosmic-ray physics. 
They are needed to calculate
 secondary anti-protons, positrons, and diffuse gamma radiation,
 which in turn provide important knowledge
 of particle propagation and the matter distribution
 in interstellar space. 
Those are also indispensable
 for studying atmospheric neutrinos. 
Although measurement of the proton and helium energy spectra
 has been performed in various experiments,
 their resultant absolute fluxes show discrepancies
 up to a factor of two at 50~GeV. 
This ambiguity causes large uncertainty in calculations of
 the atmospheric neutrinos,
 as well as secondary anti-protons, positrons, and diffuse gamma-rays. 

We report
 a new precise measurement of the cosmic-ray proton and helium spectra
 over the energy ranges of 1 to 120~GeV for protons
 and 1 to 54~GeV/nucleon for helium nuclei,
 based on a half of the data
 from a BESS balloon flight in 1998. 
The covered energy range is relevant to the atmospheric neutrinos observed as
 ``fully contained events'' in Super-Kamiokande. 
In the BESS--98 flight,
 a new trigger mode was implemented with a silica-aerogel Cherenkov counter
 to record all energetic particles
 instead of sampling the protons at a ratio of 1/60 
 as done in the previous BESS flights. 
This drastically improved statistics
 in the high-energy region above 6~GeV/nucleon,
 as reported here. 

\section{THE BESS SPECTROMETER}

\begin{figure}
  \plotone{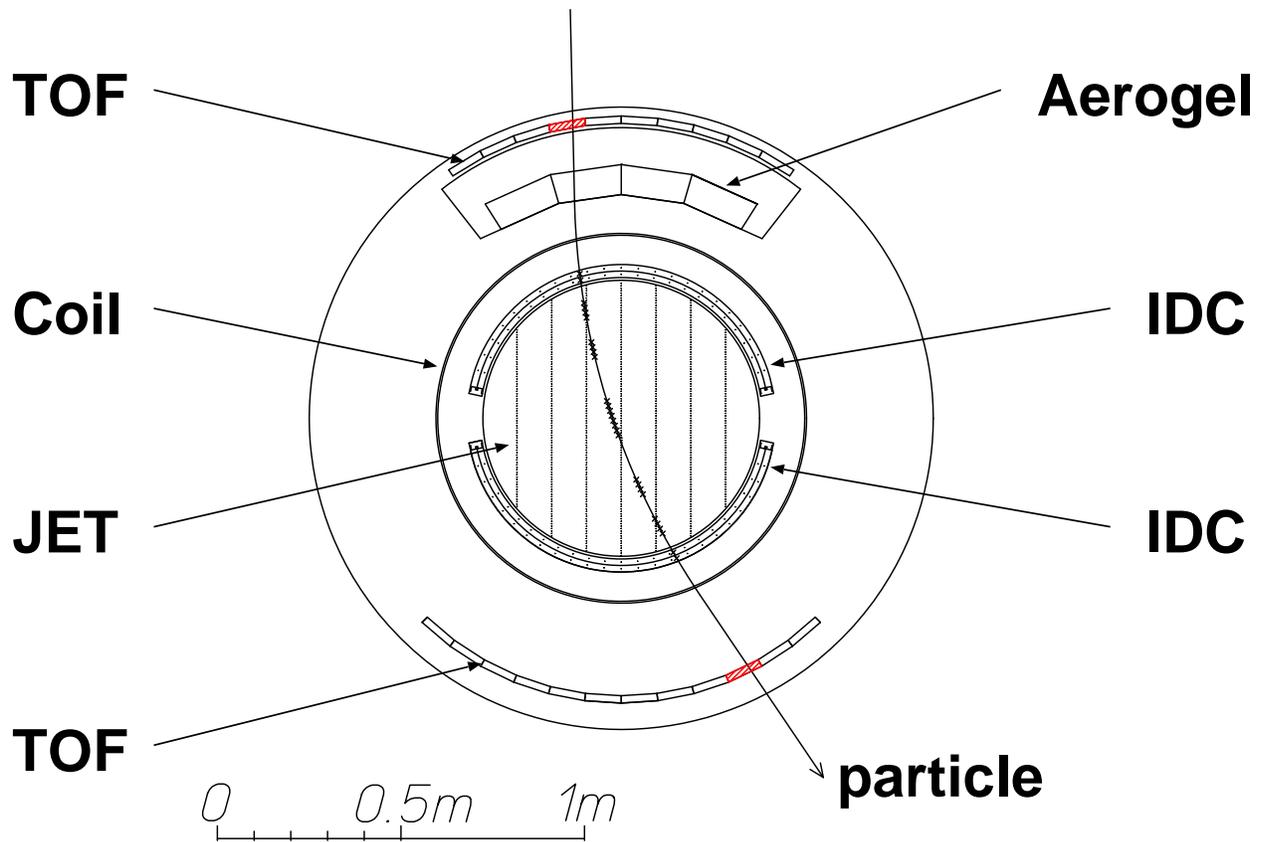}
  \caption{Cross-sectional view of the BESS instrument.
    \label{fig:besscross}}
\end{figure}

\begin{figure}
  \plotone{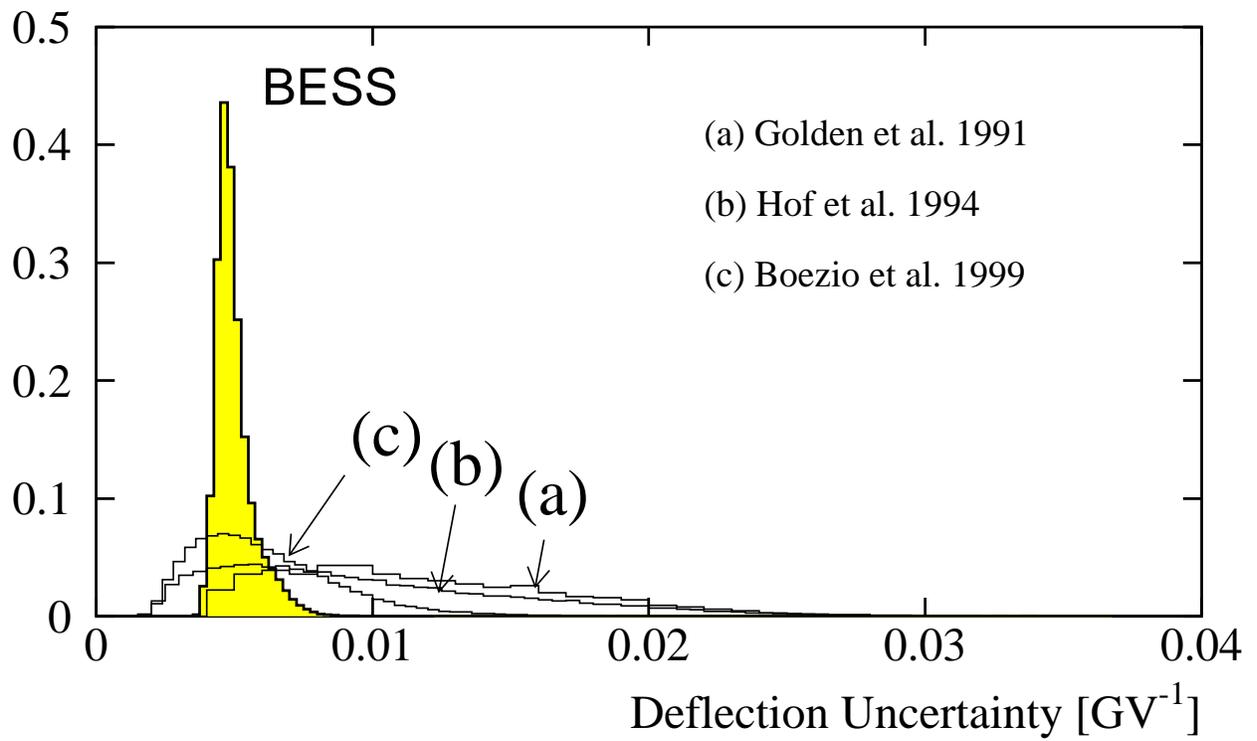}
  \caption{Deflection uncertainty for protons. 
        Each area of the histogram is normalized to unity. 
    \label{fig:defresop}}
\end{figure}

The BESS detector is
 a high-resolution spectrometer with a large acceptance
 to perform highly sensitive searches
 for rare cosmic-ray components,
 as well as precise measurement of the absolute fluxes
 of various cosmic-ray particles \citep{orito,yamamoto1994,detector}. 
As shown in Figure~\ref{fig:besscross},
 all detector components are arranged in a simple cylindrical configuration
 with a thin superconducting solenoidal magnet. 
In the central region,
 the solenoid provides
 a uniform magnetic field of 1 Tesla ($\pm 7$~\% in a fiducial volume). 
The trajectory of an incoming charged particle is measured
 by using the tracking system
 which consists of a central JET chamber and two inner drift chambers (IDC's)
 in a volume of 0.84~m in diameter and 1~m in length. 
The magnetic-rigidity ($R \equiv pc/Ze$) is reliably determined
 by a simple circular-fitting of the deflection ($R^{-1}$)
 using up-to 28 hit points
 each with a spatial resolution of 200~$\mu$m . 
Figure~\ref{fig:defresop} shows the deflection
 uncertainty for protons
 evaluated in the track fitting procedure.
Since the magnetic field is highly uniform,
 it has a narrow and sharp peak around 0.005,
 which corresponds to maximum detectable rigidity (MDR) of 200~GV,
 and has a small tail. 
The outermost detector is a set of time-of-flight (TOF) hodoscopes
 with 2~cm thick plastic scintillators. 
It provides the velocity ($\beta \equiv v/c$)
 and energy loss ($dE/dx$) measurements. 
The time resolution for energetic protons in each counter is 55~ps rms,
 resulting in a $\beta^{-1}$ resolution of 1.4~\%. 
The data acquisition sequence is initiated by a first-level TOF trigger
 by a simple coincidence of signals in the top and bottom scintillators,
 with the threshold level of 1/3 pulse height
 from minimum ionizing particles (MIP's). 
If the pulse height exceeds 2.5 times MIP signal,
 the TOF trigger is labeled as ``helium-trigger,''
 otherwise ``proton-trigger.'' 
The TOF trigger efficiency was evaluated to be $> 99.95$~\%
 from the muon data taken at sea level;
 thus, the systematic error caused by the TOF trigger inefficiency
 is negligibly small. 
In order to build a sample of unbiased triggers,
 one of every 60 and 25 ``proton-triggered'' and ``helium-triggered'' events,
 respectively,
 was recorded irrespective of succeeding online-selections. 

The instrument has a threshold-type Cherenkov counter
 with a silica-aerogel radiator
 just below the top TOF hodoscope \citep{agel}. 
The radiator was newly developed prior to the BESS--98 flight
 and it has a refractive index of 1.022. 
An auxiliary trigger was generated
 by a signal from the Cherenkov counter
 to record all of the high energy particles above $\sim$ 6~GeV
 without bias nor sampling.
The efficiency of the Cherenkov trigger
 for the energetic particles was determined
 to be $92.1 \pm 3.0$~\% by using the data sample of unbiased triggers. 

The BESS spectrometer was flown 1998 July 29-30
 from Lynn Lake, Manitoba, Canada. 
It floated at an altitude of 37~km
 (residual atmosphere of ${\rm 5~g/cm^2}$)
 with a cutoff rigidity of 0.5~GV or smaller.
The solar activity was close to the minimum. 

\section{DATA ANALYSIS}

In the first stage of data reduction,
 we selected events with a single track
 fully contained inside a fiducial volume
 defined by the central four columns out of eight columns
 in the JET chamber. 
This definition of the fiducial volume
 reduced the effective geometrical acceptance
 down to $\sim 1/3$ of the full acceptance,
 but it ensured the longest track-fitting
 and thus the highest resolution in the rigidity measurement.
The single-track selection eliminated rare interacting events. 
In order to verify the selection,
 events were scanned randomly
 in the unbiased trigger sample,
 and it was confirmed that
 995 out of 1,000 visually-identified single-track events
 passed this selection criteria
 and interacting events were fully eliminated. 
Thus, the track reconstruction efficiency was $99.5 \pm 0.2$~\%
 for a single-track event. 
In order to assure accuracy of the rigidity measurements,
 event quality such as $\chi^2$ in the track fitting procedure was imposed
 on the single-track events. 
The efficiency of this quality-check was estimated
 from loose cuts of the flight data
 to be $93.8 \pm 0.3$~\% and $93.0 \pm 0.9$~\%
 for protons and helium nuclei, respectively. 
It was almost constant over the whole energy range.

\begin{figure}
\plotone{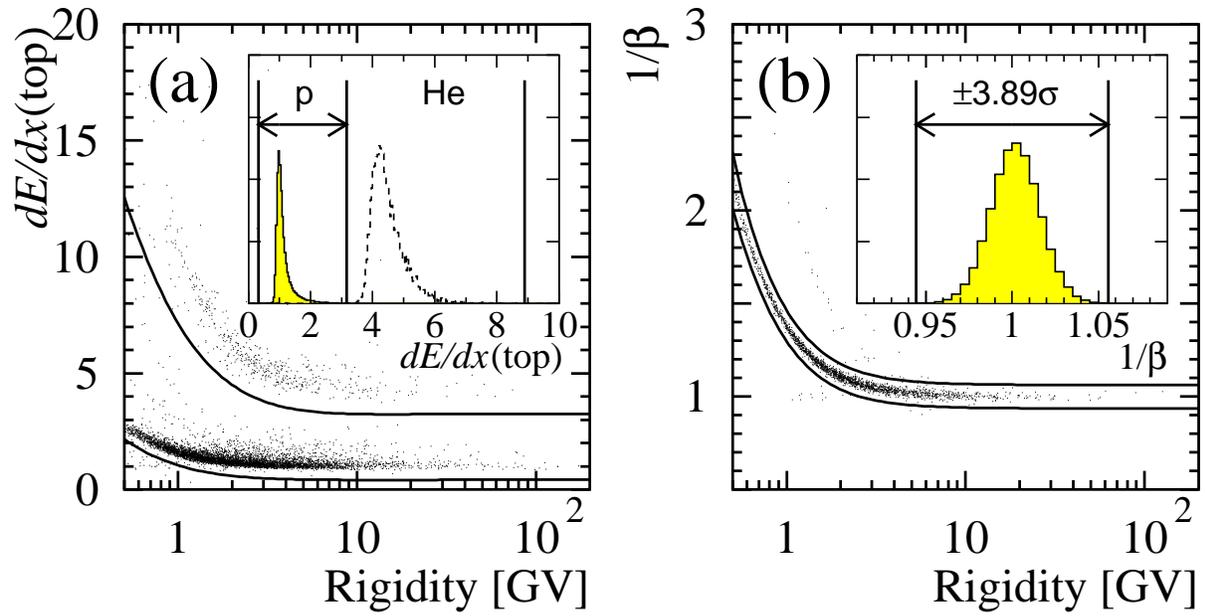}
\caption{Proton bands in (a) $dE/dx$ (top-TOF) vs rigidity plane;
 and (b) 1/$\beta$ vs rigidity plane after proton $dE/dx$ selection. 
$dE/dx$ in the bottom-TOF is also checked. 
The superimposed graphs show the proton selection criteria above 10~GV. 
Helium nuclei were selected in the same manner. 
\label{fig:pselect}}
\end{figure}

In the final stage of data reduction,
 particle identification was performed as shown in Figure~\ref{fig:pselect}
 by requiring proper $dE/dx$
 and 1/$\beta$ as a function of rigidity.
According to a study of another sample of
 $5 \times 10^5$ protons and $4 \times 10^4$ helium nuclei
 selected by using independent information
 of energy loss inside the JET chamber,
 the $dE/dx$ selection efficiencies were
 $99.3 \pm 0.2$~\% for protons and $98.2 \pm 0.7$~\% for helium nuclei,
 and the contamination probabilities should be less than 
 $3 \times 10^{-5}$ for protons
 and $4 \times 10^{-4}$ for helium nuclei. 
Since the 1/$\beta$ distribution is well described by Gaussian and
 the half width of 1/$\beta$ selection band was set at 3.89 $\sigma$,
 the efficiency is very close to unity (99.99~\% for pure Gaussian). 

Very clean proton samples were obtained below 3~GV. 
However, deuterons started to contaminate the proton band around at 3~GV,
 where the contamination was observed to be 2~\%. 
No subtraction was made for this contamination,
 because it was as small as the statistical errors
 and a deuteron-to-proton ratio decreases with energy \citep{seo1997}
 following a decrease in escape path lengths
 of primary cosmic-ray nuclei \citep{heao3}. 
In conformity with previous experiments,
 all doubly charged particle were treated as $^4$He.
With the data reduction described above,
 826,703 protons and 77,325 helium nuclei were finally identified. 
The combined efficiencies
 were
 $92.7 \pm 1.0$~\% for protons and $90.8 \pm 1.5$~\% for helium nuclei.

The geometrical acceptance defined for this analysis was calculated
 to be ${\rm 0.0851 \pm 0.0003}$~${\rm m^2sr}$
 for energetic particles
 by using simulation technique \citep{sullivan1971}. 
The simple cylindrical shape and the uniform magnetic field make it trivial
 to determine the acceptance precisely. 
The error arises from uncertainty of the detector alignment within 1~mm. 
The ratio of live data-taking-time was measured exactly to be 86.4~\%
 by counting 1~MHz clock pulses. 

The energy of each particle at the top of the atmosphere can be calculated
 by summing up the ionization energy losses
 with tracing back the event trajectory. 
Detection efficiencies were studied
 by using Monte Carlo (M.C.) simulation. 
The M.C. code was developed
 to incorporate detailed description of various interactions of helium nuclei
 into GEANT \citep{geant},
 where 
 the cross-sections and angular-distributions of the nuclear interactions
 were evaluated
 by fitting experimental data
 \citep{bizard1977,abdurakhimov1981,gasparyan1982,ableev1985,grebenyuk1989,
glagolev1993,abdullin1994}
 to energy dependent empirical formulas \citep{bradt1950}. 
The electromagnetic processes, mainly due to $\delta$--rays,
 are also treated properly. 
They are more significant in helium nuclei interactions,
 because cross-sections in electromagnetic processes ($\sigma_{em}$)
 behaves as $\sim Z^2$
 whereas those in hadronic processes ($\sigma_{had}$)
 is approximately proportional to $(2Z)^{2/3}$.
The M.C. well reproduced the observed event shape. 
The simulated and observed number of hit-counters in the bottom TOF hodoscope,
 for instance,
 were agree within a discrepancy of 0.9~\% and 1.8~\%
 for proton and helium, respectively. 
The systematic errors in the M.C. originate mainly in uncertainties
 of $\sigma_{had}$ and $\sigma_{em}$. 
We attributed relative errors of
 $\pm 5$~\% to $\sigma_{had}({\rm p+A})$,
 $\pm 5$~\% to $\sigma_{em}({\rm p+A})$,
 $\pm 15$~\% to $\sigma_{had}({\rm He+A})$,
 $\pm 20$~\% to $\sigma_{em}({\rm He+A})$, and
 $\pm 20$~\% to $\sigma_{had}({\rm CNO+A})$.
Another source of a systematic error in the M.C. was
 the uncertainty of the material distribution inside the BESS spectrometer,
 which was estimated to be $\pm 10$~\%. 
The probability that cosmic-ray protons and helium nuclei, respectively,
 can pass through the whole detector without interaction was
 $87.6 \pm 2.3$~\% and $74.5 \pm 6.8$~\% at 1~GeV,
 and $79.7 \pm 2.9$~\% and $64.6 \pm 7.5$~\% at 100~GeV. 

According to similar M.C. studies,
 the probabilities
 that primary cosmic-rays can penetrate
 the residual atmosphere of 5~g/cm$^2$
 is about 94~\% and 90~\% for the proton and helium, respectively,
 over the entire energy range discussed here.
Atmospheric secondary protons,
 which account for about 3.5~\% at 1~GeV 
 and less than 1.5~\% above 10~GeV of observed protons,
 were subtracted based on the calculation by \citet{papini}. 
Atmospheric secondary helium above 1~GeV/nucleon
 is dominated by
 fragments of heavier cosmic-ray nuclei (mainly Carbon and Oxygen). 
The flux ratio of atmospheric secondary helium to primary C+O
 was calculated to be 0.14
 at a depth of $5~{\rm g/cm^2}$,
 based on the total inelastic cross-sections of CNO + Air interactions and
 the helium multiplicity
 in ${\rm ^{12}C + CNO}$ interactions \citep{ahmad1989}. 
The total correction of atmospheric secondary helium
 due to all nuclei with $Z>2$
 was estimated
 to be about 2~\% over the entire energy range.
The ambiguity in this estimation
 arises from the uncertainties in $\sigma_{had}({\rm CNO+A})$
 and in absolute fluxes of heavier cosmic-ray nuclei,
 to which we attributed relative error of $\pm 20$~\%. 
The systematic errors originating in the correction of the residual air effect
 were estimated to be  $\pm 0.3$~\% for protons
 and $\pm 2.0$~\% helium nuclei.

Because of the finite resolution in rigidity measurement,
 and the very steep spectral shape,
 the observed spectrum shape may suffer deformation. 
The effect of finite resolution was estimated by simulation,
 where the error in rigidity measurement was tuned
 to reproduce the distribution shown in Figure~\ref{fig:defresop}.
The effect was found to be smaller than 1~\% below 25~GV,
 but it became visible with increasing rigidity. 
The observed spectrum gradually gets lower
 than original spectrum, with a ratio of $- 2.5$~\% at 70~GV
 and then rapidly rises to $\pm 0$~\% around 120~GV. 
No correction was made for this deformation,
 because the effect is as small as the statistical errors
 over the energy range discussed here.

\section{EXPERIMENTAL RESULTS AND CONCLUSION}

\begin{figure}
\plotone{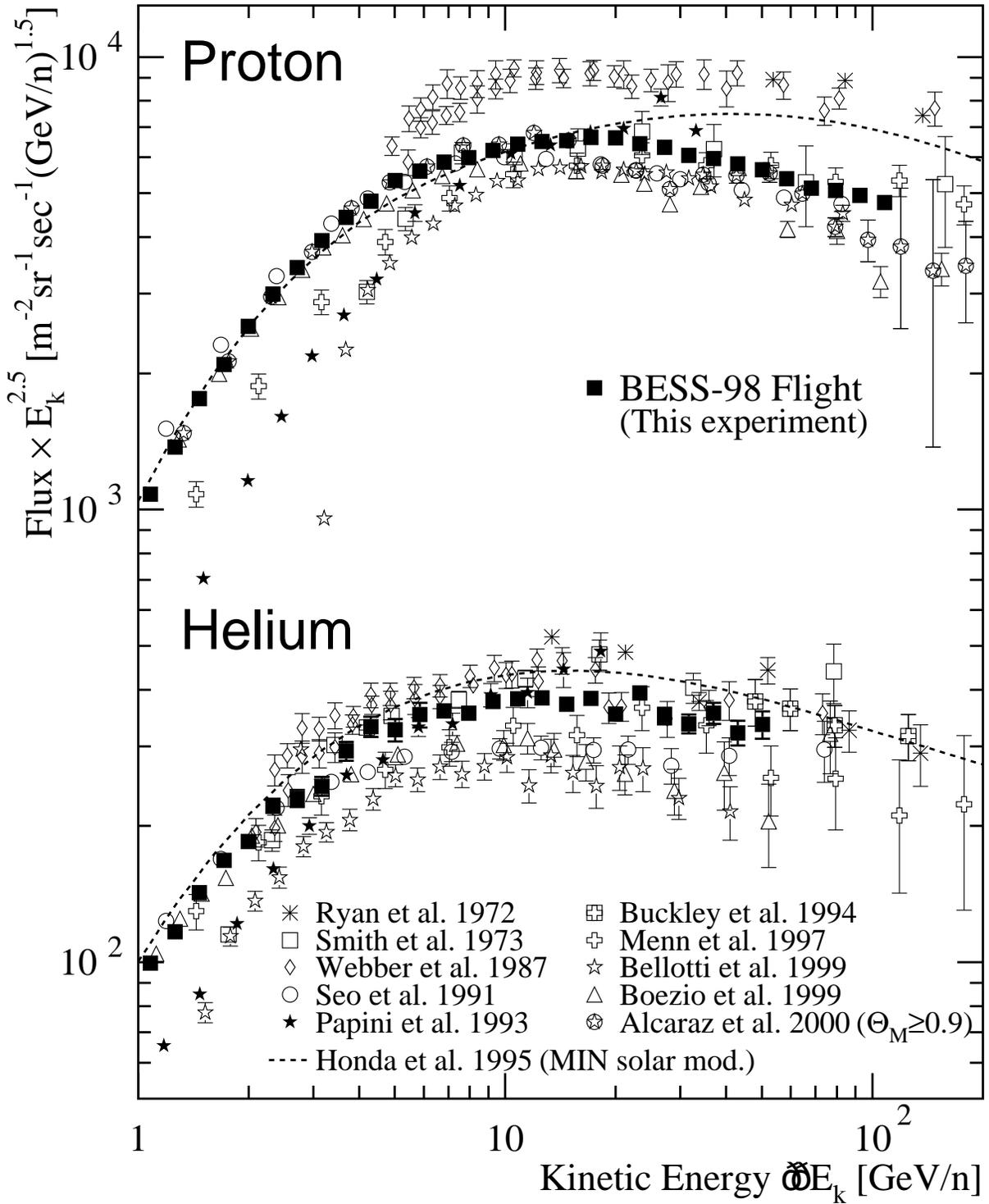}
\caption{Absolute differential proton and helium spectra. 
  Filled squares show results of the BESS--98 experiment. 
  The spectra obtained by other experiments are also shown
  by different symbols indicated in the figure.
  Dashed lines show assumed spectra
  in the atmospheric neutrino flux calculation. 
  \label{fig:pftoa}}
\end{figure}

The proton and helium fluxes at the top of the atmosphere have been
 obtained from the BESS--98 flight data
 as summarized in Table~\ref{tbl:result}
 and as shown in Figure~\ref{fig:pftoa}
 in comparison with the previous experiments. 
The first and second errors in Table~\ref{tbl:result} represent
 statistical and systematic errors, respectively. 
The overall errors including both errors
 are less than $\pm 5$~\% for protons
 and $\pm 10$~\% for helium nuclei. 
The dotted lines in Figure~\ref{fig:pftoa} indicate the spectra
 assumed in the calculation of atmospheric neutrinos \citep{hkkm}.
Our results, as well as other recent measurements,
 are favorable to lower fluxes than
 the ones assumed in the atmospheric neutrino calculation
 especially above a few tens of GeV. 
It may suggest an importance of the reconsideration
 for the atmospheric neutrino flux predictions. 
Precise measurements of primary cosmic-rays
 will help to improve the accuracy
 in the atmospheric neutrino calculations.

\acknowledgments

The authors would thank NASA and NSBF for the balloon flight operation. 
This experiment was supported
 by Grants-in-Aid from Monbusho
 and Heiwa Nakajima Foundation in Japan 
 and by NASA in the U.S.A. 
The analysis was performed with the computing facilities
 at ICEPP, University of Tokyo.

\clearpage

\begin{deluxetable}{rrlccllccl}
\tabletypesize{\footnotesize}
\tablewidth{0pc} 
\tablecolumns{10} 
\tablecaption{Proton and helium fluxes at the top of the atmosphere.
 \label{tbl:result}} 
\tablehead{
 \colhead{} & \colhead{} & \colhead{}
  & \multicolumn{3}{c}{Proton} & \colhead{} & \multicolumn{3}{c}{Helium} \\
 \cline{4-6} \cline{8-10} \\
 \multicolumn{2}{c}{Energy Range} & \colhead{}
  & \colhead{$\overline{E_k}$}
  & \multicolumn{2}{l}{${\rm Flux\pm\Delta Flux_{sta}\pm\Delta Flux_{sys}}$}
  & \colhead{}
  & \colhead{$\overline{E_k}$}
  & \multicolumn{2}{r}{${\rm Flux\pm\Delta Flux_{sta}\pm\Delta Flux_{sys}}$} \\
 \multicolumn{2}{c}{[GeV/n]} & \colhead{}
  & \colhead{[GeV/n]}
  & \multicolumn{2}{c}{[${\rm m^{-2}sr^{-1}sec^{-1}(GeV/n)^{-1}}$]}
  & \colhead{}
  & \colhead{[GeV/n]}
  & \multicolumn{2}{c}{[${\rm m^{-2}sr^{-1}sec^{-1}(GeV/n)^{-1}}$]}
}
\startdata 
  1.00 & 1.17 & & 1.08 & 8.92 $\pm$ 0.12 $\pm$ 0.22 & $\times 10^{2} $ & & 1.08 & 8.21 $\pm$ 0.26 $\pm$ 0.62 & $\times 10     $ \\ 
  1.17 & 1.36 & & 1.26 & 7.72 $\pm$ 0.11 $\pm$ 0.19 & $\times 10^{2} $ & & 1.26 & 6.57 $\pm$ 0.22 $\pm$ 0.50 & $\times 10     $ \\ 
  1.36 & 1.58 & & 1.47 & 6.74 $\pm$ 0.09 $\pm$ 0.17 & $\times 10^{2} $ & & 1.47 & 5.46 $\pm$ 0.19 $\pm$ 0.41 & $\times 10     $ \\ 
  1.58 & 1.85 & & 1.71 & 5.46 $\pm$ 0.08 $\pm$ 0.14 & $\times 10^{2} $ & & 1.71 & 4.38 $\pm$ 0.15 $\pm$ 0.33 & $\times 10     $ \\ 
  1.85 & 2.15 & & 2.00 & 4.52 $\pm$ 0.07 $\pm$ 0.11 & $\times 10^{2} $ & & 2.00 & 3.29 $\pm$ 0.12 $\pm$ 0.25 & $\times 10     $ \\ 
  2.15 & 2.51 & & 2.33 & 3.63 $\pm$ 0.05 $\pm$ 0.09 & $\times 10^{2} $ & & 2.33 & 2.69 $\pm$ 0.10 $\pm$ 0.21 & $\times 10     $ \\ 
  2.51 & 2.93 & & 2.71 & 2.83 $\pm$ 0.04 $\pm$ 0.07 & $\times 10^{2} $ & & 2.71 & 1.90 $\pm$ 0.08 $\pm$ 0.15 & $\times 10     $ \\ 
  2.93 & 3.41 & & 3.16 & 2.22 $\pm$ 0.04 $\pm$ 0.06 & $\times 10^{2} $ & & 3.15 & 1.38 $\pm$ 0.07 $\pm$ 0.11 & $\times 10     $ \\ 
  3.41 & 3.98 & & 3.68 & 1.71 $\pm$ 0.03 $\pm$ 0.05 & $\times 10^{2} $ & & 3.68 & 1.12 $\pm$ 0.05 $\pm$ 0.09 & $\times 10     $ \\ 
  3.98 & 4.64 & & 4.30 & 1.27 $\pm$ 0.02 $\pm$ 0.03 & $\times 10^{2} $ & & 4.30 & 8.65 $\pm$ 0.44 $\pm$ 0.66 & $              $ \\ 
  4.64 & 5.41 & & 5.01 & 9.65 $\pm$ 0.19 $\pm$ 0.26 & $\times 10     $ & & 5.01 & 5.80 $\pm$ 0.34 $\pm$ 0.44 & $              $ \\ 
  5.41 & 6.31 & & 5.84 & 6.89 $\pm$ 0.15 $\pm$ 0.19 & $\times 10     $ & & 5.84 & 4.27 $\pm$ 0.27 $\pm$ 0.33 & $              $ \\ 
  6.31 & 7.36 & & 6.81 & 4.91 $\pm$ 0.02 $\pm$ 0.20 & $\times 10     $ & & 6.80 & 2.96 $\pm$ 0.04 $\pm$ 0.24 & $              $ \\ 
  7.36 & 8.58 & & 7.94 & 3.43 $\pm$ 0.01 $\pm$ 0.14 & $\times 10     $ & & 7.92 & 1.99 $\pm$ 0.03 $\pm$ 0.16 & $              $ \\ 
  8.58 & 10.0 & & 9.25 & 2.42 $\pm$ 0.01 $\pm$ 0.10 & $\times 10     $ & & 9.24 & 1.44 $\pm$ 0.03 $\pm$ 0.12 & $              $ \\ 
  10.0 & 11.7 & & 10.8 & 1.70 $\pm$ 0.01 $\pm$ 0.07 & $\times 10     $ & & 10.8 & 9.98 $\pm$ 0.20 $\pm$ 0.81 & $\times 10^{-1}$ \\ 
  11.7 & 13.6 & & 12.6 & 1.18 $\pm$ 0.01 $\pm$ 0.05 & $\times 10     $ & & 12.6 & 6.82 $\pm$ 0.15 $\pm$ 0.55 & $\times 10^{-1}$ \\ 
  13.6 & 15.8 & & 14.6 & 8.05 $\pm$ 0.04 $\pm$ 0.33 & $              $ & & 14.6 & 4.50 $\pm$ 0.12 $\pm$ 0.36 & $\times 10^{-1}$ \\ 
  15.8 & 18.5 & & 17.1 & 5.57 $\pm$ 0.03 $\pm$ 0.23 & $              $ & & 17.1 & 3.16 $\pm$ 0.09 $\pm$ 0.26 & $\times 10^{-1}$ \\ 
  18.5 & 21.5 & & 19.9 & 3.78 $\pm$ 0.03 $\pm$ 0.16 & $              $ & & 19.9 & 1.99 $\pm$ 0.07 $\pm$ 0.16 & $\times 10^{-1}$ \\ 
  21.5 & 25.1 & & 23.2 & 2.51 $\pm$ 0.02 $\pm$ 0.10 & $              $ & & 23.2 & 1.51 $\pm$ 0.05 $\pm$ 0.12 & $\times 10^{-1}$ \\ 
  25.1 & 29.3 & & 27.1 & 1.67 $\pm$ 0.01 $\pm$ 0.07 & $              $ & & 27.0 & 9.15 $\pm$ 0.39 $\pm$ 0.74 & $\times 10^{-2}$ \\ 
  29.3 & 34.1 & & 31.5 & 1.10 $\pm$ 0.01 $\pm$ 0.05 & $              $ & & 31.5 & 5.98 $\pm$ 0.29 $\pm$ 0.49 & $\times 10^{-2}$ \\ 
  34.1 & 39.8 & & 36.8 & 7.35 $\pm$ 0.08 $\pm$ 0.31 & $\times 10^{-1}$ & & 36.9 & 4.30 $\pm$ 0.23 $\pm$ 0.35 & $\times 10^{-2}$ \\ 
  39.8 & 46.4 & & 42.9 & 4.87 $\pm$ 0.06 $\pm$ 0.20 & $\times 10^{-1}$ & & 42.9 & 2.65 $\pm$ 0.17 $\pm$ 0.22 & $\times 10^{-2}$ \\ 
  46.4 & 54.1 & & 50.0 & 3.22 $\pm$ 0.05 $\pm$ 0.14 & $\times 10^{-1}$ & & 49.8 & 1.88 $\pm$ 0.13 $\pm$ 0.16 & $\times 10^{-2}$ \\ 
  54.1 & 63.1 & & 58.3 & 2.10 $\pm$ 0.04 $\pm$ 0.09 & $\times 10^{-1}$ & & \nodata & \multicolumn{2}{c}{\nodata}                \\ 
  63.1 & 73.6 & & 67.9 & 1.36 $\pm$ 0.03 $\pm$ 0.06 & $\times 10^{-1}$ & & \nodata & \multicolumn{2}{c}{\nodata}                \\ 
  73.6 & 85.8 & & 79.3 & 9.17 $\pm$ 0.20 $\pm$ 0.39 & $\times 10^{-2}$ & & \nodata & \multicolumn{2}{c}{\nodata}                \\ 
  85.8 & 100. & & 92.6 & 6.08 $\pm$ 0.15 $\pm$ 0.26 & $\times 10^{-2}$ & & \nodata & \multicolumn{2}{c}{\nodata}                \\ 
  100. & 117. & & 108. & 4.00 $\pm$ 0.12 $\pm$ 0.17 & $\times 10^{-2}$ & & \nodata & \multicolumn{2}{c}{\nodata}                \\ 
\enddata 
\end{deluxetable}

\clearpage

\end{document}